\documentclass{mem}

\usepackage{natbib}
\usepackage{txfonts}
\usepackage{graphicx}

\begin{document}

\title{Simbol-X capability of detecting the non-thermal emission of stellar flares}
\subtitle{}

\author{C.~Argiroffi\inst{1, 2} \and G.~Micela\inst{2} \and A.~Maggio\inst{2}}
\offprints{C.~Argiroffi}
\institute{Dip. di Scienze Fisiche ed Astronomiche, Universit\`a di Palermo, Piazza del Parlamento 1, 90134 Palermo, Italy, \email{argi@astropa.unipa.it} \and INAF - Osservatorio Astronomico di Palermo, Piazza del Parlamento 1, 90134 Palermo, Italy, \email{giusi@astropa.unipa.it, maggio@astropa.unipa.it }}

\titlerunning{Detecting the non-thermal emission of stellar flares}
\authorrunning{C.~Argiroffi et al.}

\abstract
{We investigate the capability of detecting, with Simbol-X, non-thermal emission during stellar flares, and distinguishing it from hot thermal emission. We find that flare non-thermal emission is detectable when at least $\sim20$\,cts are detected with the CZT detector in the $20-80$\,keV band. Therefore Simbol-X will detect the non-thermal emission from some of the X-ray brightest nearby stars, whether the thermal vs. non-thermal relation, derived for solar flares, holds.}

\maketitle{}

\section{Introduction}

Stellar flares are phenomena where the magnetic field rapidly releases large amount of energy. It is supposed that: 1) magnetic reconnections generate non-maxwellian population of fast particles; 2) these particles hit and heat the chromospheric plasma, evaporating it, and producing non-thermal emission (hard X-rays and $\gamma$-rays); 3) the evaporated chromospheric plasma fills coronal structures and generates thermal emission (mostly in the UV and soft X-ray bands).

In solar data a power-law spectrum, compatible to the bremsstrahlung emission of fast particles hitting a thick target, is observed \citep[i.e.][]{HudsonRyan1995}. This hard power-law emission precedes the soft X-ray and UV thermal emission. Information on stellar emission in the hard X-ray band is very poor, since most X-ray observations are dedicated only to the soft band. Recently \citet{OstenDrake2007} presented the detection of non-thermal emission in the hard X-rays during a stellar flare.

The study of the flare non-thermal emission allows to study: 1) the depicted scenario for stellar flares, and hence whether and how the thermal emission is caused by the non-thermal particles; 2) the flare energy balance, since only a small amount of released energy is lost by thermal radiation.

We investigate the Simbol-X capability of detecting non-thermal emission associated with stellar flares, and distinguishing it from a hot thermal one. Results about this issue are important also for other astrophysical subjects where non-thermal emission is likely present (galaxy clusters, SNR).

\section{Method}

We adopt the following procedure: 1) we simulate MPD and CZT spectra adopting as model the superposition of a thermal and a non-thermal component; 2) we assume as null hypothesis that the simulated spectra can be described by thermal components only; hence we fit simultaneously MPD and CZT spectra adopting a 2-$T$ optically thin plasma as model; 3) we assume that the null hypothesis of only thermal emission is rejected, and hence non-thermal emission is detected and recognized, when the {\it EM} of the hottest component is significantly greater than zero and its temperature is larger than 300\,MK.

To produce simulated MPD and CZT spectra we consider typical flares of nearby active stars. We are not interested in time resolved spectroscopy, hence we consider all the parameters as time-averaged over the flare duration. In all cases we consider a flare duration of 20\,ks. For each star we produce different simulated spectra: we keep frozen the thermal component, and vary the normalization of the non-thermal component. With this choice for a fixed thermal emission component we are able to check the amount of non-thermal emission needed to detect it.

\begin{figure}[]
\begin{center}
\resizebox{0.95\hsize}{!}{\includegraphics[clip=true]{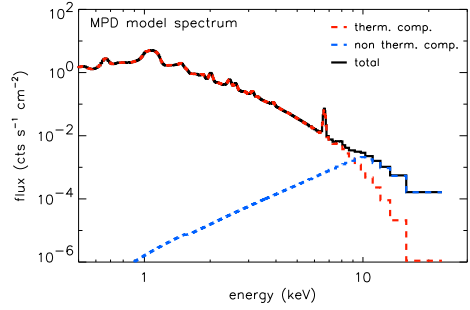}}
\end{center}
\caption{
\footnotesize One of the model adopted for simulating flare spectra.}
\label{fig:mod}
\end{figure}

We use the {\tt mekal} XSPEC model to simulate the thermal component (Table~\ref{tab:par} contains the parameters of the thermal emission for the selected stellar targets). We simulate the non-thermal emission using the broken power law XSPEC model ({\tt bknpower}). We assume: a photon index of 3 for $E<E_{0}$, in order to have a negligible contribution of the non-thermal component at low energy; a break point $E_{0}=10$\,keV; a photon index of -2 for $E>E_{0}$. Figure~\ref{fig:mod} shows one of the explored models.

\begin{table}
\scriptsize
\caption{Model parameters}
\label{tab:par}
\begin{tabular}{lcccc}
\hline
Name             & $D$  & {\it EM}                   & $T$    & $\log L_{XT}^{a}$ \\
                 &      &                            &        &               \\
                 & (pc) & ($10^{50}\,{\rm cm^{-3}}$) & (MK)   & (erg/s)       \\
\hline
Prox Cen         & 1.3  & 5                          & 20     & 27.5 \\
AD Leo           & 4.7  & 10                         & 20     & 27.7 \\
EV Lac           & 5.0  & $2\times10^2$              & 20     & 28.9 \\
AB Dor           & 14.9 & $2\times10^2$              & 20     & 29.0 \\
HR 1099          & 29.0 & $1\times10^3$              & 20     & 29.6 \\
PMS stars        & 150  & $5\times10^4$              & 20     & 31.3 \\
\hline
\multicolumn{5}{l}{$^{a} L_{XT}$ is computed in the $1.55-12.4$\,keV band.} \\
\end{tabular}
\end{table}

\begin{figure}[]
\begin{center}
\resizebox{0.95\hsize}{!}{\includegraphics[clip=true]{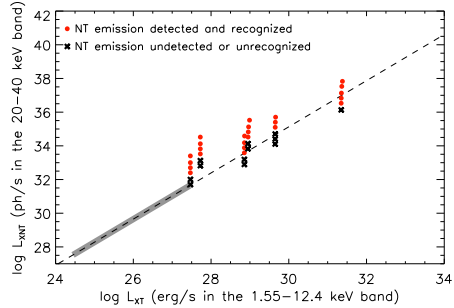}}
\end{center}
\caption{\footnotesize Luminosity of the thermal and non-thermal component for the flare sample analyzed. Gray area marks the solar flare range.}
\label{fig:res}
\end{figure}

\section{Results and conclusions}

Inspecting all the simulated flares we find that: non-thermal emission is detected and recognized with Simbol-X observations whether the CZT detector collects more than $\sim20$ photons in the $20-80$\,keV band.

For each simulated spectrum we compute the thermal and non-thermal luminosity. We compare the results obtained with those derived from solar flares by \citet[][see Fig.~5]{IsolaFavata2007}. We find that for some of the considered flares the non-thermal emission can be detected with Simbol-X if thermal vs. non-thermal solar flare relation is valid also at high luminosities.

\bibliographystyle{aa} 
\bibliography{argiroffi}

\begin{thebibliography}{3}
\expandafter\ifx\csname natexlab\endcsname\relax\def\natexlab#1{#1}\fi

\bibitem[{{Hudson} \& {Ryan}(1995)}]{HudsonRyan1995}
{Hudson}, H. \& {Ryan}, J. 1995, ARA\&A, 33, 239

\bibitem[{{Isola} {et~al.}(2007){Isola}, {Favata}, {Micela}, \&
  {Hudson}}]{IsolaFavata2007}
{Isola}, C., {Favata}, F., {Micela}, G., \& {Hudson}, H.~S. 2007, \aap, 472,
  261

\bibitem[{{Osten} {et~al.}(2007){Osten}, {Drake}, {Tueller}, {Cummings},
  {Perri}, {Moretti}, \& {Covino}}]{OstenDrake2007}
{Osten}, R.~A., {Drake}, S., {Tueller}, J., {et~al.} 2007, ApJ, 654, 1052

\end{thebibliography}

\end{document}